\documentclass[9pt,twocolumn,twoside]{osajnl}
\journal{ol}
\setboolean{shortarticle}{true}

\usepackage{braket}
\usepackage{gensymb}
\usepackage{adjustbox}
\usepackage[justification=justified]{caption}
\usepackage[labelformat=empty,skip=0pt]{subcaption}
\usepackage{hyperref}

\newcommand*\phantomsubfigure[1]{\begin{subfigure}{0pt}\caption{}\label{#1}\end{subfigure}\ignorespaces}%

\title{Zeeman-tunable Modulation Transfer Spectroscopy}

\author[1,*]{Chloe So}
\author[1]{Nicholas L. R. Spong}
\author[1,2]{Charles M\"{o}hl}
\author[1,3]{Yuechun Jiao}
\author[1]{Teodora Ilieva}
\author[1]{Charles S. Adams}

\affil[1]{Joint Quantum Centre (Durham-Newcastle), Department of Physics, Durham University, South Road, Durham, DH1 3LE, United Kingdom.}
\affil[2]{Physikalisches Institut, Universit\"{a}t Heidelberg, Im Neuenheimer Feld 226, 69120 Heidelberg, Germany.}
\affil[3]{State Key Laboratory of Quantum Optics and Quantum Optics Devices, Institute of Laser Spectroscopy, Shanxi University, Taiyuan 030006, China.}

\affil[*]{Corresponding author: chloe.so@durham.ac.uk}

\begin{abstract}
    Active frequency stabilization of a laser to an atomic or molecular resonance underpins many modern-day AMO physics experiments. With a flat background and high signal-to-noise ratio, modulation transfer spectroscopy (MTS) offers an accurate and stable method for laser locking. Despite its benefits, however, the four-wave mixing process that is inherent to the MTS technique entails that the strongest modulation transfer signals are only observed for closed transitions, excluding MTS from numerous applications. Here, we report for the first time the observation of a magnetically tunable MTS error signal. Using a simple two-magnet arrangement, we show that the error signal for the $^{87}$Rb $F=2 \rightarrow F'=3$ cooling transition can be Zeeman-shifted over a range of $>$15 GHz to any arbitrary point on the rubidium $\text{D}_2$ spectrum. Modulation transfer signals for locking to the $^{87}$Rb $F=1 \rightarrow F'=2$ repumping transition as well as 1 GHz red-detuned to the cooling transition are presented to demonstrate the versatility of this technique, which can readily be extended to the locking of Raman and lattice lasers.
\end{abstract}

\setboolean{displaycopyright}{true}

\begin{document}

\maketitle
\section{Introduction}
    Stabilization of a laser against frequency fluctuations, or laser `locking', is a key technique in many areas of research. This is especially relevant in the field of atomic and molecular physics, where often the laser has to be stabilized well below the natural linewidth of an atomic transition, often to less than a few hundred kilohertz. To this end, a number of spectroscopic techniques have been developed, including single-beam methods such as dichroic atomic vapor laser locking (DAVLL) \cite{corwin_frequency-stabilized_1998,millett-sikking_davll_2006,mccarron_heated_2007}; and pump-probe schemes such as saturation absorption spectroscopy (SAS) \cite{preston_dopplerfree_1996}, polarization spectroscopy (PS) \cite{wieman_doppler-free_1976,pearman_polarization_2002}, frequency modulation spectroscopy (FMS) \cite{bjorklund_frequency-modulation_1980}, and modulation transfer spectroscopy (MTS) \cite{shirley_modulation_1982,zhang_characteristics_2003,mccarron_modulation_2008}. Of these approaches, FMS and MTS require external modulation of the laser, followed by coherent demodulation of the probe, to generate the dispersive lock signals. The main differences are that FMS relies on the direct detection of vapor absorption and dispersion by the modulated probe beam, whereas the MTS lineshape originates from the frequency-dependent four-wave mixing (FWM) process \cite{raj_high-frequency_1980,ducloy_theory_1982} which \textit{transfers} modulation from the pump to the probe.
    
    \begin{figure*}
        \centering
        \includegraphics[width=.88\textwidth]{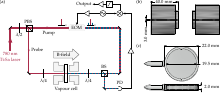}%
            \phantomsubfigure{fig:fig1a}
            \phantomsubfigure{fig:fig1b}
            \phantomsubfigure{fig:fig1c}
        \caption{(a) Schematic of the Zeeman-tunable modulation transfer spectroscopy experimental setup (${\lambda}/2$ = half-wave plate, ${\lambda}/4$ = quarter-wave plate, BS = 50:50 non-polarizing beamsplitter, EOM = electro-optic modulator, PBS = polarizing beamsplitter, PD = photodiode). The sideband-modulated pump and unmodulated probe beams are allowed to propagate collinearly through a heated rubidium cell of natural abundance and length 2 mm, across which a uniform magnetic field of 0.6 T is applied. The electronic signal lines are drawn here in black. (b) The N52 grade neodynium magnets and (c) 2 mm Rb vapor cell used in the experiment.}
        \label{fig:fig1}
    \end{figure*}

\begin{figure}
        \centering
        \includegraphics{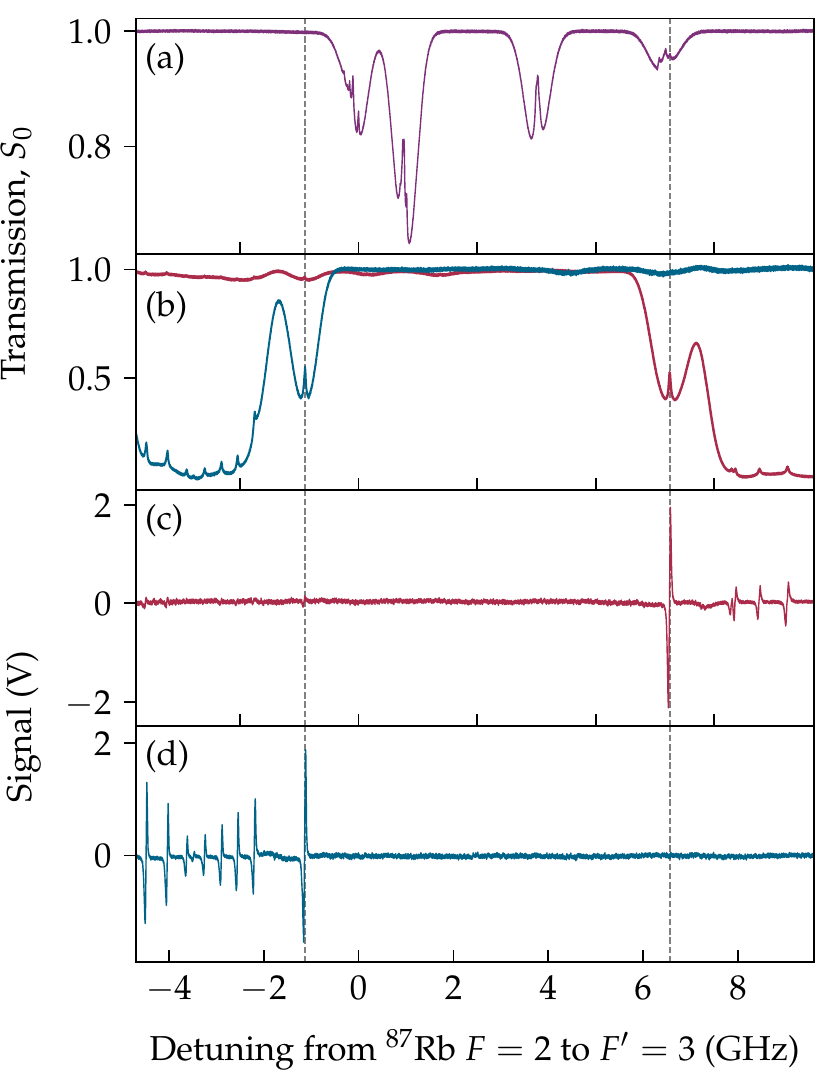}
        \caption{Doppler-free spectrum of the $\text{D}_2$ line in Rb vapor with (a) $B=0$ and (b) 0.6 T; and the Zeeman-shifted MTS spectrum using (c) right- and (d) left-circularly polarized light. (c) shows the MTS error signal for the $^{87}$Rb $F=2 \rightarrow F'=3$ cooling transition Zeeman-shifted onto the $^{87}$Rb $F=1 \rightarrow F'=2$ repumping transition. (c) and (d) are interchangeable by a $90 \degree$-rotation of the quarter-wave plates. Color online: red, with RCP incident light; blue, with LCP incident light.}
        \label{fig:fig2}
    \end{figure}

    In particular, MTS has emerged as one of the most commonly used laser locking techniques. The reason for this is twofold: firstly, the method readily creates dispersive-like lineshapes that reside on a symmetric, Doppler-free background (see Figs.~\ref{fig:fig2} and~\ref{fig:fig3}). This is because the phase-matching criterion of FWM is not satisfied away from resonance. The lineshape baseline stability thus becomes insensitive to the residual linear-absorption of the medium, and is immune, unlike FMS, to dispersive elements (e.g. parasitic etalons) which can alter the location of the lock-point by adding an offset to the demodulated error signal. The zero-crossings of the modulation transfer signals are accurately centered on the corresponding hyperfine peaks, providing stable, unambiguous frequency references to which the laser can be locked.

    Another reason to prefer MTS-based laser locking is that the generated signal is suppressed for open transitions and undesirable crossover features arising from the pump-probe scheme. This is as a result of the FWM process only being efficient for closed transitions, where dissipation to atomic states other than the ground state is minimal and the interaction time between atom and light is long \cite{zhe_li_modulation_2011,noh_modulation_2011}. The resultant spectrum is `clean', displaying strong signals with steep gradients for closed transition lines only. Whilst this can be useful in the event that the frequency spacing between consecutive transitions is small, as is usually the case in the alkali metals, it also presents a problem when one wishes to lock \textit{away} from the closed transition---an example of this is the $^{87}$Rb $F=1 \rightarrow F'=2$ repumping transition. FMS is advantageous in this regard, as it is able to produce error signals for all sub-Doppler features with amplitudes corresponding to those in the saturation absorption spectrum. Once again, however, the non-zero background that accompanies the FMS technique means that the lock point is at risk of `hopping' (amongst other pathologies) from one transition to another.

    In answer to the above, and following growing interest in performing thermal vapor experiments in the hyperfine Paschen--Back (HPB) regime \cite{olsen_optical_2011,sargsyan_hyperfine_2014,sargsyan_complete_2015,sargsyan_atomic_2015,sargsyan_hyperfine_2018,whiting_electromagnetically_2015,whiting_direct_2016,whiting_single-photon_2017,whiting_four-wave_2018,mathew_simultaneous_2018}, we present in this Letter a technique to shift the MTS error signal in the presence of a large magnetic field. This method exploits the Zeeman effect, wherein the spectral line of an atom can be split in the presence of an external magnetic field; as well as the fact that so long as a transition is closed we obtain a modulation transfer lineshape, to produce multiple `copies' of lock signals that are magnetically tunable. Specifically, we will show that it is possible to Zeeman-shift the MTS signal for the $^{87}$Rb $F=2 \rightarrow F'=3$ cooling transition by $\pm8$ GHz, including onto the $^{87}$Rb $F=1 \rightarrow F'=2$ repumping transition, using a 0.6 T magnetic field.
    
    It is worth mentioning that, while the application of bias fields to spectroscopy setups is not new, previous demonstrations lack either the high signal-to-noise ratio of the MTS technique \cite{marchant_off-resonance_2011, reed_low-drift_2018}, or the flexibility of having a widely-tunable stabilized source that is the focus of this paper \cite{zi_laser_2017,long_magnetic-enhanced_2018}. This contrasts with the proposed method which, aside from preserving all the aforenarrated advantages of MTS, has the additional benefits of being highly adaptable and reproducible.
    
\section{Experimental Demonstration}
    
    The experimental setup for Zeeman-tunable MTS (ZMTS) is illustrated in Fig.~\ref{fig:fig1a}. The experiment uses an ultra-narrow linewidth MSquared SolsTiS laser to achieve the large spectral range (15 GHz), but typical diode lasers may also be used if a smaller range is desired. 4.6 mW of 780 nm light is separated into pump (3.5 mW; $1/\text{e}^2$ radius $0.52\pm0.01$ mm) and probe (1.1 mW; $1/\text{e}^2$ radius $0.55\pm0.01$ mm) beams with a low-order half-wave plate and a polarizing beamsplitter. The pump beam is phase modulated by an electro-optic modulator (Photonic Technologies EOM-02-12.5-V), driven below the resonant frequency at 8.5 MHz to prevent `kinking' of the error signal \cite{mccarron_modulation_2008}. A non-polarizing beamsplitter is used to reflect the pump and its accompanying sidebands into a 2 mm long vapor cell of natural abundance rubidium (Fig.~\ref{fig:fig1c}), where it interacts with the counterpropagating probe beam via the $\chi^3$ susceptibility of the medium in a FWM process. The generated sideband, now induced onto the probe, beats with the carrier to produce an oscillating signal at the modulation frequency $\omega_m$. This signal is detected on a fast photodiode (Hamamatsu C10508-01), amplified (Mini-Circuits ZFL-500), downmixed (Mini-Circuits ZX05-1L-S), and finally fed into a proportion-integration-differentiation (PID) controller (Toptica FALC 110) to output the MTS error signal. The relative phase shift between the modulation signal and the reference signal supplied to the mixer can be set digitally using a two-channel RF generator (Tektronix AFG 1062) with phased-matched outputs and variable offset. Note also that because of its small size, the vapor cell is heated to 120 $\degree$C (Thorlabs HT19R) to maintain a sufficient optical depth for generating the error signals of Fig.~\ref{fig:fig2}.

    \begin{figure}
        \centering
        \includegraphics{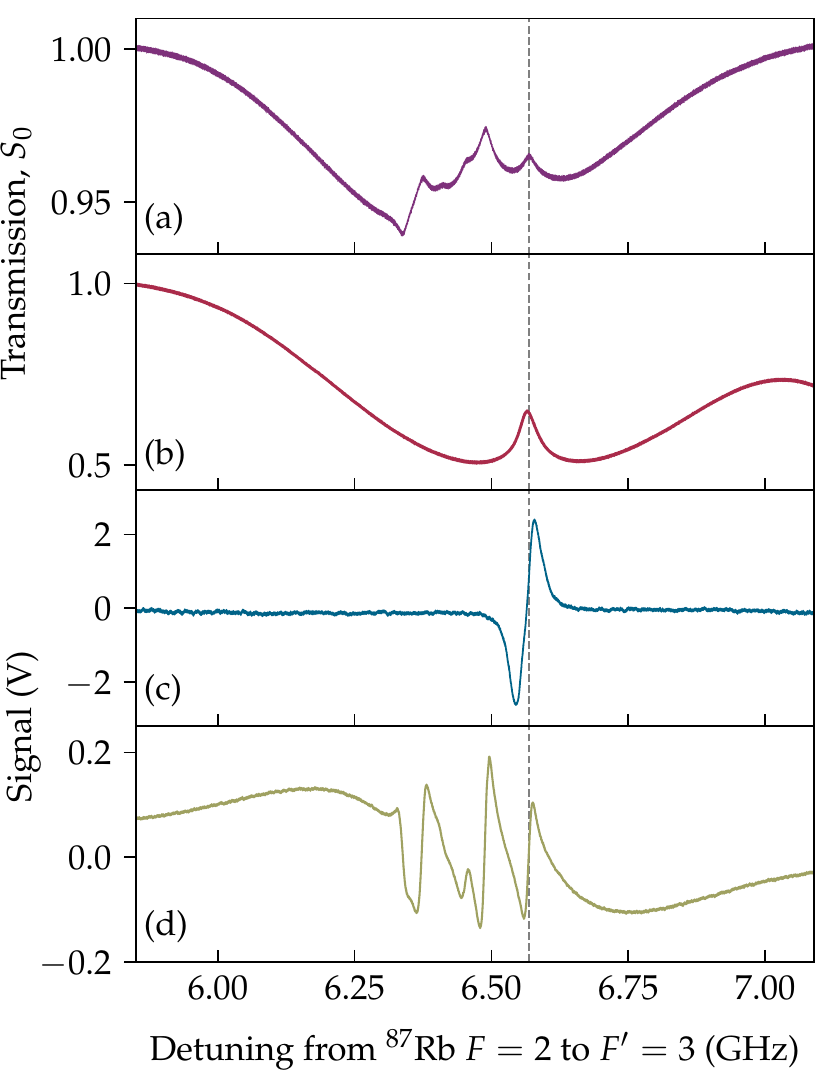}
        \caption{(a) Normal and (b) Zeeman-shifted pump-probe spectroscopy for the $^{87}$Rb $F=1 \rightarrow F'$ transitions. Note that the dip at the $F=1 \rightarrow F'=0$ transition (leftmost in the first panel) is caused by optical pumping. A comparison between (c) Zeeman-shifted MTS and (d) FMS for the $^{87}$Rb $F=1 \rightarrow F'=2$ repumping transition.}
        \label{fig:fig3}
    \end{figure}

    This setup differs from conventional MTS in that a large constant magnetic field is applied to the vapor cell along the axis of the propagating beams. This is provided by a pair of `top-hat' N52 NdFeB magnets (custom-made by and purchased from \href{http://www.jinmagnets.com/en/index/index.asp}{Shanghai Jinmagnets}), as depicted in Fig.~\ref{fig:fig1b} in a cross-sectional view. The magnetic field strength, and thus the position of the lock signal, can be varied easily by changing the separation between the magnets. In our case, field strengths of up to 0.6 T that are homogeneous over the length of the cell (rms variation 4 \micro T) are permitted by the small size of the cell and relatively powerful magnets. Unshielded, the current setup has a leakage field that falls to that of the Earth's at a distance of 0.7 m---further information regarding field uniformity and magnet design can be found in Refs.~\cite{whiting_nonlinear_2017,zentile_applications_2015}. The two quarter-wave plates on either side of the magnets serve to control the circularity, and more importantly the handedness, of the incoming pump and probe polarizations. This allows the selective driving of $\sigma^+$ and $\sigma^-$ transitions with right- (RCP) and left-circularly polarized (LCP) light respectively, adding another level of tunability to the already versatile method. In conjunction with the applied magnetic field, this supports up to $>$15 GHz of freedom in the error signal location (in practice, anywhere on the rubidium $\text{D}_2$ spectrum). To illustrate, Fig.~\ref{fig:fig2} shows that by changing only the incident polarization of the pump and probe one may choose from two lock points that are several gigahertz apart; whereas Fig.~\ref{fig:fig3} displays an MTS signal translated magnetically by 6.6 GHz from the $^{87}$Rb $F=2 \rightarrow F'=3$ cooling transition onto the $^{87}$Rb $F=1 \rightarrow F'=2$ repumping transition.
    
     \begin{figure}
        \centering
        \includegraphics{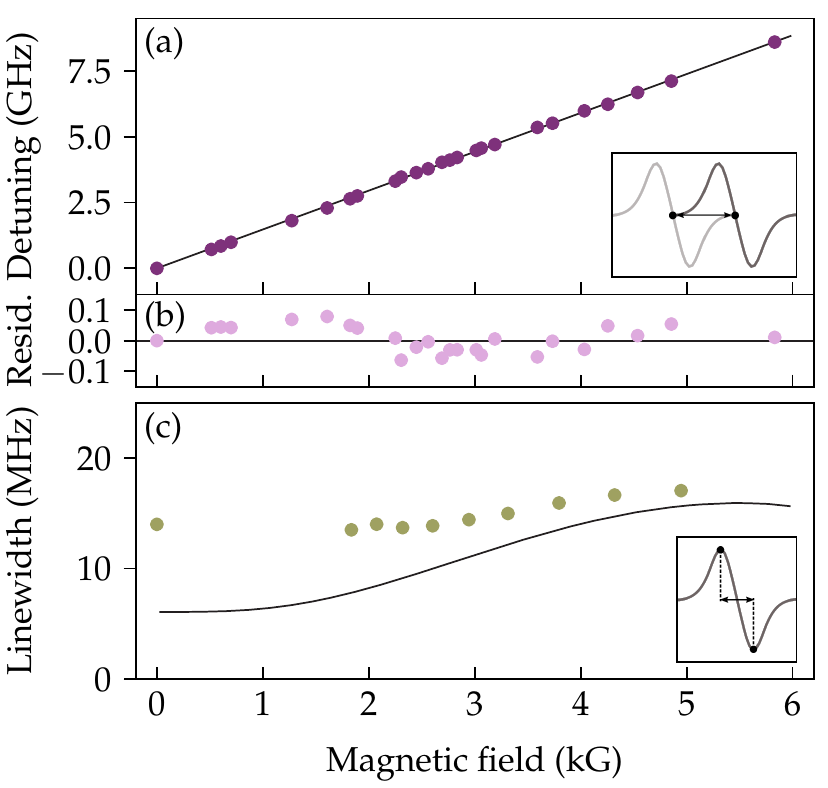}
            \phantomsubfigure{fig:fig4a}
            \phantomsubfigure{fig:fig4b}
            \phantomsubfigure{fig:fig4c}
        \caption{A plot of (a) detuning and (c) linewidth of the $^{87}$Rb $F=2 \rightarrow F'=3$ cooling transition as a function of magnetic field strength. The solid black lines in (a) and (c) indicate respectively the theoretically predicted Zeeman-shift of the $\ket{F=2, m_F=2} \rightarrow \ket{F'=3, m_{F}'=3}$ transition, and \textit{ab initio} calculations of its linewidth based on the $B$-field homogeneity within the cell. The error bars in both plots are too small to be visible. (b) displays the normalized residuals of (a), with the discrepancies arising due to non-linearities in the laser scan. Insets: schematics indicating how the experimental shift and linewidth are determined.}
        \label{fig:fig4}
    \end{figure}
    
        The effect of the external magnetic field on the detuning and linewidth of an atomic transition was investigated. As in Figs.~\ref{fig:fig4a} and ~\ref{fig:fig4b}, the experimental detuning shows excellent agreement with the theoretically predicted Zeeman-shift for the $^{87}$Rb $\ket{F=2, m_F=2} \rightarrow \ket{F'=3, m_{F}'=3}$ transition, enabling one to anticipate the frequency shift simply by knowing the magnetic field. Fairly good agreement is also found between the observed and calculated linewidths at high magnetic field strengths (Fig.~\ref{fig:fig4c}). Here, the theoretical linewidth is obtained by taking into account the Zeeman-broadening arising from the $B$-field variation within the cell. The breakdown of the model at weaker fields is suggestive of the fact that alternative broadening mechanisms---namely, power \cite{zienau_optical_1975, milonni_lasers_1988} and transit-time broadening \cite{borde_saturated_1976,thomas_transit-time_1977,rautian_saturation_1970,baklanov_transit_1975}---are dominant and in play.
    
    \begin{figure}
        \centering
        \includegraphics{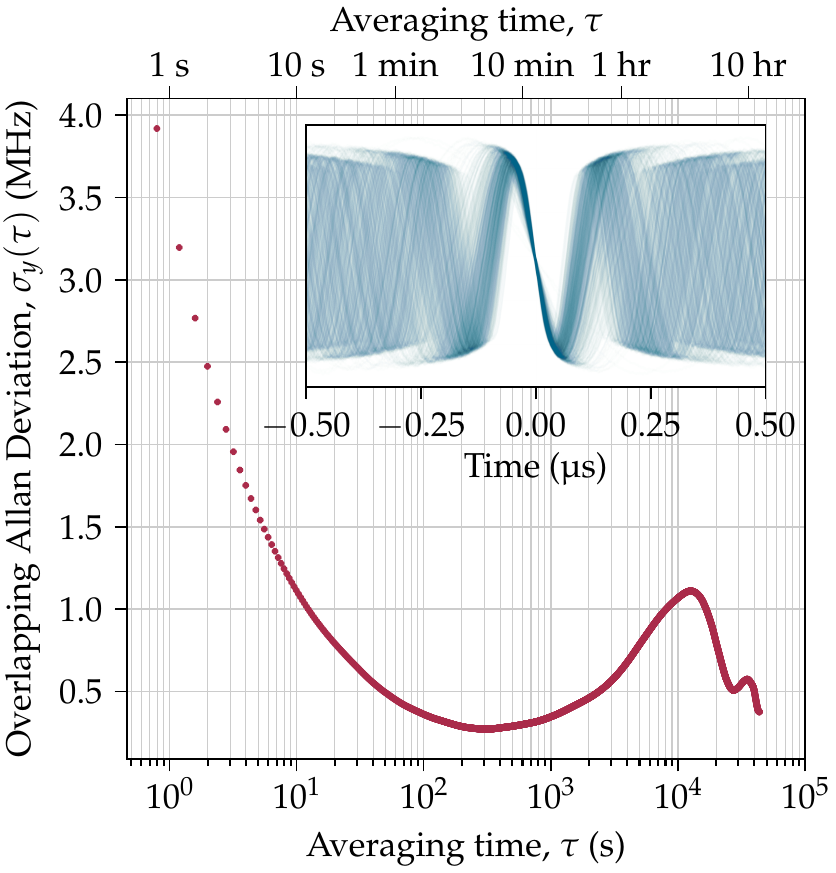}
        \caption{Overlapping Allan deviation of the beat note frequency between the locked laser and a separate, MTS-stabilized laser. Taken at 7 MHz detuning, the inset shows a persistence-type plot of the beat signal averaged over 1000 shots.}
        \label{fig:fig5}
    \end{figure}
    
    To evaluate the long-term stability of the method, the overlapping Allan deviation \cite{riley_handbook_2008,allan_statistics_1966} of the beat note frequency between the locked laser (Toptica DL Pro) and an independent, MTS-stabilized laser was measured over a period of 24 hours. The laser remains locked for the entirety of this duration and exhibits frequency fluctuations well below the natural linewidth of the atomic transition (6 MHz), as can be seen from Fig.~\ref{fig:fig5} which shows a \textit{combined} frequency instability of <1 MHz for the two lasers over times up to $10^3$ s. The longer-term deprecation ($\tau>10^4$ s) is attributed to drifts in ambient temperature. The combined coherence time of the two lasers, which is the FWHM of the approximately Gaussian envelope of the beat, is found to be 0.33 \micro s.
    
    The demonstrated stability is sufficiently good for various applications such as laser cooling and, if required, could be further improved by incorporating the intensity stabilisation method discussed in Ref.~\cite{zi_laser_2017}.
    
\section{Conclusion}

    In conclusion, we have presented in this Letter a technique to arbitrarily translate a modulation transfer signal on the rubidium $\text{D}_2$ spectrum by virtue of a large axial magnetic field. Locking the laser to the shifted signal gives a frequency stability of better than 1 MHz for timescales up to $10^3$ s. Though the technique was demonstrated in the context of the $^{87}$Rb $\text{D}_2$ line, the same method can be conveniently extended to other atomic species and types of lasers.
    
    \bigskip
    
    \noindent Funding. Engineering and Physical Sciences Research Council (EPSRC) (EP/M014398/1).
    
    \bigskip
    
    \noindent Acknowledgements. The design files for all relevant parts and figure data are available on the Durham University Collections repository (\href{http://dx.doi.org/10.15128/r2r207tp335}{doi:10.15128/r2r207tp335}).

    \bigskip

\bibliography{references}
\bibliographyfullrefs{references}

\end{document}